**Emergence of large-scale mechanical spiral waves in bacterial living matter**

Shiqi Liu [1], Ye Li[1], Yuhao Wang[1], Yilin Wu [1*]

[1] *Department of Physics and Shenzhen Research Institute, The Chinese University of Hong Kong, Shatin, NT, Hong Kong, P.R. China.*

*To whom correspondence should be addressed. Mailing address: Room 306, Science Centre, Department of Physics, The Chinese University of Hong Kong, Shatin, NT, Hong Kong, P.R. China. Tel: (852) 39436354. Fax: (852) 26035204.
Email: ylwu@cuhk.edu.hk

**Abstract**:
Propagating spiral waves have been discovered in various chemical, biological and physical systems. Spiral waves in multicellular organisms are often associated with essential living functions. Although certain eukaryotic microorganisms have long been known to generate spiral waves, evidence of spiral wave pattern has been lacking in the bacterial world. Here we report the discovery of a unique form of propagating spiral waves in dense bacterial populations where cells engage in cyclic force-generating processes driven by type-IV pilus motility. Specifically, we discovered that synchronization of pilus activity in the bacterial living matter leads to large-scale spatiotemporal regulation of tension force in the form of propagating spiral waves. Theoretical modelling reveals that the spiral tension waves result from nonreciprocity in cell-cell interactions. Our findings reveal a novel mechanism of large-scale force regulation in bacterial world and may shed light on the emergent mechanics of biofilms and microbiomes. Pilus-driven bacterial living matter also provides a mechanical active medium for studying electrical or chemical spiral waves in living systems.

## Introduction

Propagating waves with spiral patterns are discovered in various chemical, biological and physical systems (1-3) such as chemical excitable medium (4-6), cardiac tissue (7), and neural networks (8). The onset of spiral waves in multicellular organisms is often associated with essential living functions (8, 9). For example, spiral waves of electrochemical activities in neural tissues serve as a rhythmic organizer in cortex neurons (8), while those in cardiac tissues may cause ventricular arrhythmia (7, 10). Although certain eukaryotic microorganisms have long been known to generate spiral waves (11, 12), spiral wave patterns have not been observed in the bacterial world.

Here we report the discovery of a unique form of propagating spiral waves in dense bacterial populations where cells engage in cyclic force-generating processes driven by type IV pili. Type IV pilus is a motile organelle shared by diverse bacterial species (13, 14) and it is the most powerful molecular motor known to date (15). We discovered that synchronization of pilus activity in the bacterial living matter leads to large-scale spatiotemporal regulation of mechanical forces in the form of propagating spiral tension waves. The spiral tension wave is highly stable with stationary cores, resembling those seen in the electrical activity of heart tissues during ventricular arrhythmia (7, 10). Theoretical modelling suggests that the striking pattern of spiral tension waves is a result of nonreciprocal coupling between pilus activities. As type IV pili are widespread in bacteria, our findings may shed light on the emergent mechanics of biofilms (16) and microbiomes in natural and clinical settings. Moreover, the unique wave pattern provides a tractable mechanical analogue for studying electrical or chemical spiral waves in diverse living systems (7, 10, 17, 18).

## Results

### Stable propagating spiral waves emerge in bacterial living matter via a self-organized synchronization process

Powered by an ATP-driven translational molecular motor (13, 14), type IV pilus extends and retracts in a cyclic manner (19-22), generating pulling forces up to ~100 pN (15). Using naturally-developed surface colonies of the model organism *Pseudomonas aeruginosa* (14), we discovered the presence of propagating spiral waves in *P. aeruginosa* colonies. The waves appeared in the form of ordered light-intensity oscillation in the phase-contrast images of the colonies over a macroscopic length scale much greater than the cell size (~2-4 μm in length and 0.8 μm in width) (Fig. 1A,B and Movie 1). The period and the wavefront propagating speed of such spiral waves ranged from ~3-10 min and ~1-3 μm/s, respectively. Type IV pilus motor activity is essential for the wave generation, since *P. aeruginosa* strains with defect in type IV pilus motility [either not producing type IV pili (PA14 *flgk::Tn5 ΔpilA*) or not being able to retract and generate mechanical forces (PA14 *flgk::Tn5 ΔpilY1*); Methods] failed to form the propagating spiral waves.

To follow the onset and evolution of the propagating spiral waves, we prepared disk-shaped artificial bacterial films ~20-30 μm in thickness and densely packed with a homogeneous population of *P. aeruginosa* cells (Methods). The solid-like artificial bacterial film also displays propagating spiral waves, and more importantly, it allows us to exclude the effect of behavioral heterogeneity in naturally developed bacterial colonies. Using this experimental system, we found a self-organization process during the formation of the spiral wave. In the beginning, some sporadic domains with segmented and disordered wavefronts appeared in the artificial bacterial films (Fig. 1C, upper; Movie 2). These disordered wavefronts spontaneously emerged; subsequently they coalesced, inter-connected, and eventually turned into propagating spiral waves. To understand the spatiotemporal ordering process of the local light-intensity oscillations, we defined the instantaneous period of the wave at a specific location as the time interval between two consecutive wavefronts passing this location, and used this information to compute the phase of local oscillations (setting the phase at the wavefronts seen in phase-contrast images as zero) (Methods). The phase distribution allows us to calculate the local order parameter $\chi(\vec{r})$ commonly adopted to describe synchronization behavior and defined as $\chi(\vec{r})e^{i\bar{\theta}} = \frac{1}{N}\sum_s e^{i\theta}$ (2, 23), where $\bar{\theta}$ is the local average of phase angle ($\theta$) of oscillators within a local circular area $s$ centered at $\vec{r}$ (Methods). As shown by distributions of the phase and the local order parameter (Fig. 1C, middle and lower; Movie 2), the phases of local light-intensity oscillations self-organized in space from a disordered state to a highly ordered state with the spiral wave pattern.

Moreover, after the wave form stabilized, the spiral cores (or spiral centers) specified by the local minima of local order parameter (6) did not display any meandering motion seen in many other systems with propagating spiral waves (24); instead, they appeared to be stationary (Movie 2), with a negligible core diffusivity as low as $6.92$ µm$^2$/min or equivalently $4.80 \times 10^{-5}$ square wavelengths per period (wavelength ~0.99 mm) (Fig 1D; Fig. S1,S2; Methods). We further calculated instantaneous oscillation period at all positions during the spiral wave development and found that the spatial distribution of period gradually homogenized during the emergence of the spiral waves (Fig. 1E; Fig. S3, lower row). To summarize, the ordering of phase distribution and the homogenization of local oscillation period together demonstrate a synchronization process that leads to formation of the propagating spiral waves with remarkable temporal stability.

**The propagating spiral waves manifest spatiotemporal ordering of tension force in bacterial living matter**

The observed light-intensity oscillation in the phase-contrast images presumably reflects the variation of surface-packing cell density in the quais-2D bacterial living matter. To examine this notion, we labeled *P. aeruginosa* cells with the fluorescent protein mCherry and measured cell density via fluorescence imaging (Methods). We found that the mCherry fluorescence intensity (a proxy of local cell density) exhibits the same propagating spiral wave pattern that coincides with the wave observed in phase-contrast images (Movie 3). Cell density at the leading side of the wavefront is higher than that at the lagging side (Fig. 2A; Movie 3); also during one wave period, the cell density at a specific location first increases as a wavefront approaches and then drops abruptly as the wavefront moves away (Fig. 2A). The observed propagating spiral waves thus reflect periodic variation of cell density.

To understand the origin of the cell density variation in the bacterial film, we tracked the movement of cells by labeling a small fraction of the population (0.1%) with mCherry fluorescent protein (Methods). We found that cells appeared to displace gradually toward the wave propagating direction together with the approaching wavefronts for a distance of several µm, and then they rapidly returned to their original position after the wavefronts had passed (Fig. 2B, left; Movie 4). These results suggest that the cell density variation is associated with periodic forth-and-back displacement of cells that causes in-plane deformation of the bacterial film. Therefore, the wave resembles a longitudinal or acoustic wave seen in passive elastic materials; here cells serve as an active medium that not only mediates wave propagation but also self-generates the wave through pilus motor activity. Interestingly, while the displacement of cells lagged behind the density variation (Fig. 2B, left), the time rate of cell displacement was well synchronized with the density variation (Fig. 2B, right). Our theoretical analysis of displacement field shows that this result is a direct consequence of mass conservation in the bacterial film (Methods).

The local forth-and-back displacement of cells associated with the cell density variation displayed negligible net displacement during a cycle, suggesting local force imbalance in the bacterial film. To analyze the forces in the bacterial film, we modelled the bacterial film as an elastic medium in which the elastic force and substrate friction competes with an internal tension that arises from the pulling forces generated by pilus retraction (Methods). We found that the variation of tension force inside the bacterial film is in phase with the variation in local cell density (Methods). In other words, the observed spiral waves of density variation shown in Fig. 2A are in fact propagating tension waves in the bacterial film, with cells at the leading side of the wavefront experiencing a higher pilus-generated tension than the lagging side. This pattern of propagating tension waves would require spatiotemporal coordination of pilus motor activity in mass elements of the bacterial film. For instance, the average pilus retraction probability should be higher at the leading side and lower at the lagging side of the wavefront. In particular, the rapid restoring displacement of cells when wavefronts pass by indicates that pilus motors have by then completed retraction synchronously.

**The spiral tension waves are driven by synchronized bacterial pilus activity**

To examine whether coordination of bacterial pilus motor activity underlies the propagating tension waves, we sought to manipulate the waves by treating the bacterial films with drugs that alter pilus motor activity. EGTA is a calcium chelator that was previously reported to cause pilus retraction (25, 26). When a filter paper disc soaked with EGTA (Methods) was placed onto a bacterial film prior to the emergence of segmented and disordered wavefronts (as shown in Fig. 1C), we found that a train of waves emanated from near the edge of the filter paper disc and propagated outwards, which displayed similar local dynamics of cell density variation and forth-and-back motion pattern as seen in the naturally emerged propagating tension waves (Fig. 2C, Movie 5). In bacterial films with pre-existing propagating tension waves, the EGTA treatment abolished the original wave pattern near the filter disk, followed by the emergence of a similar wave train initiated near the edge of the filter disk (Movie 5). In addition, we found that filter disks containing other drugs that were reported to cause pilus retraction (trifluoperazine and thioridazine (26)) also triggered train waves (Movie 6). Interestingly, a diffusion pulse of extracellular ATP (eATP) showed a similar effect in exciting train waves (Movie 6); eATP was reported to inhibit pilus motility and may also reset the phase of the cyclic force-generating process, although it is unclear whether it causes pilus retraction or inhibits pilus extension (27).

These results can be understood as follows. Cells near the filter disk experience a rapid diffusion pulse of the drug (e.g., EGTA), and the type IV pili of cells that encounter the drug concentration pulse will simultaneously retract. The simultaneous pilus retraction would reset the phase of the cyclic force-generating process of pili and force the synchronization of these cycles, thus triggering the train waves. Therefore, these results provide strong evidence that the propagating tension waves results from spatiotemporal coordination of pilus motor activity. The coordination or coupling of pilus activities is presumably mediated by the mechanosensing pathways of cells, as type IV pilus motility is actively coupled to the mechanical environment (28-33).

**A model of locally coupled oscillators with nonreciprocal interaction reproduces the observed propagating spiral waves**

Next we seek to understand how spatiotemporal coordination of pilus retraction-extension cycle could give rise to the propagating spiral tension waves. Although methods to visualize type IV pili have been developed for isolated cells (20-22, 34), direct observation of pilus activity in densely packed cell layers is technically challenging. Here we resort to mathematical modeling and adopt the framework of Kuramoto model (2) well-known for the study of synchronization phenomena. We considered mass elements in the bacterial film as a system of spatially coupled oscillators on a two-dimensional square lattice (Fig. 3A). Each oscillator describes the pilus retraction-extension cycle represented by a phase angle $\theta$ ($\theta \in [0, 2\pi]$), which is proportional to the probability that the pilus motors of cells in a mass element entering the retraction or force-generating state; for instance, $\theta = 0$ corresponds to the state that none of the pilus motors is retracting and thus the mass element does not exert pilus-mediated forces to the environment. As the pilus retraction-extension cycle can only proceed forward but not backward (35), a key ingredient of our model is that an oscillator is more likely to accelerate than to decelerate its phase during interaction with its neighbours, i.e. the bacterial pilus activities are coupled in a nonreciprocal manner. Specifically, we introduce isotropic nonreciprocal coupling to the locally coupled two-dimensional Kuramoto model as follows (Fig. 3A; Methods):

$$\frac{d\theta_{(i,j)}}{dt} = \omega_{0(i,j)} + \frac{\kappa}{N}\sum_{r \leq l}[\sin(\theta_{(m,n)} - \theta_{(i,j)}) \cdot F(\sin(\theta_{(m,n)} - \theta_{(i,j)}))] + \eta(t). \quad \#(1)$$

Here $\theta_{(i,j)}$ is the phase angle of the oscillator at the location $(i,j)$; $\omega_{0(i,j)}$ is the intrinsic angular frequency of the oscillator located at $(i,j)$; $r = \sqrt{(m-i)^2 + (n-j)^2}$ is the spatial distance between two oscillators at point $(m,n)$ and $(i,j)$; $l$ denotes the coupling range; $\kappa$ is the coupling strength; and $\eta(t)$ is a Gaussian white noise (see Methods). The step function $F(x)$ in the coupling term is defined as: $F(x) = (1-\varepsilon)/2$, for $x < 0$; $F(x) = (1+\varepsilon)/2$, for $x \geq 0$, with $\varepsilon \in [0,1]$ being a measure of nonreciprocity. This is the simplest form of isotropic, nonreciprocal coupling (Methods).

In addition to the Kuramoto dynamics for the phase oscillators described above, we allow the mass elements to be displaced due to external forces (including pilus-driven tension force, elastic force, and friction force; Methods). The displacement of mass elements changes the relative distance between neighbouring phase oscillators and thereby may modify the coupling dynamics between these locally-coupled oscillators. Taking the tension $T$ generated by a mass element to be proportional to the phase angle $\theta$ associated with the mass element, the two-dimensional displacement field $\vec{d}(x,y,t) = (d_1, d_2)$ of the bacterial film can be solved numerically and the cell density distribution $\rho(x,y,t)$ can be obtained by $\rho \propto -\nabla \vec{d}$ (Methods).

We performed numerical simulations of the nonreciprocal coupled-oscillator model based on Eq. 1 on a finite homogeneous domain (Methods) with $\varepsilon > 0$. Starting from initially random distributions of phase angle and intrinsic angular frequency (Methods), we found that the system self-organizes into a highly ordered state that displays periodically propagating spiral waves at any level of nonreciprocity. For the case with full nonreciprocity ($\varepsilon = 1$, which better describes the coupling of bacterial pilus activities), the evolutionary process of the spiral wave pattern faithfully reproduces the dynamics observed in experiments (Fig. 3B-E; Movie 7), including the spatial homogenization of instantaneous angular frequency (Fig. S4) and the stability of spiral cores (with a negligible spiral core diffusivity of $6.14 \times 10^{-4}$ square wavelengths per period; Fig. 3F, Fig. S2). By contrast, without nonreciprocal coupling (i.e., setting $\varepsilon = 0$), Eq. 1 becomes the familiar Kuramoto model that does not spontaneously generate stable spiral waves in a homogeneous system. We note that another modified Kuramoto model could produce propagating spiral waves by introducing a time delay in the coupling function (36); however, the spiral wave there was less stable, with the spiral cores displaying apparent motion. We further found that forced synchronization of oscillators in certain area triggers traveling train waves that emanate from the area's boundary and eliminates pre-existing spiral waves outside the area (Fig. S5, Movie 8; Methods). The results reproduce the phenomena when artificial bacterial films were treated by drugs causing pilus retraction (Movie 5, Movie 6), and are similar to the unpinning process of spiral waves previously reported in B-Z reactions and cardiac tissues (37). Taken together, the remarkable agreement between our simulations and experiments supports that nonreciprocal coupling of pilus activities is key to the emergence of the observed propagating spiral waves.

**Stability of the spiral wave pattern depends on nonreciprocity and is resilient to spatial inhomogeneity**

An important feature of the propagating spiral waves found in our experiments is that the spiral cores are stationary, which is reproduced by our simulations described above. To further understand the stability of spiral cores, we followed the evolutionary dynamics of spiral waves in simulations. We found that the number density of spiral cores increases with the level of nonreciprocity; at lower nonreciprocity, nearby pairs of spiral cores with opposite topological charges (+1 or -1 for counterclockwise or clockwise rotating spiral waves) tend to annihilate at a higher rate during the developmental process, thus leaving less cores when the spiral wave pattern stabilizes (Fig. 4A; Movie 9). To understand this behavior, we would like to study pairwise interaction between spiral cores systematically in the model. We note that the dynamics of spiral cores in the model is barely affected by oscillator displacement; this observation can be understood as arising from the fact that the displacement of mass elements (a few micron; Fig. 3E) is 1-2 orders of magnitude smaller than the coupling range set in the model (160 μm, which is required to reproduce the wavelength of spiral wave patterns found in experiments; Methods). Therefore, for the sake of generality, we now fix the position of oscillators in the model and study spiral core dynamics with the nonreciprocal Kuramoto model (Eq. 1) alone.

We artificially create two counter-rotating Archimedean spiral wave patterns (i.e. the equi-phase lines of the oscillators being described by $\rho \propto \phi$, where $\rho$ and $\phi$ are the radius and polar angle in polar coordinates, respectively) (3) with their cores separated by a variable distance; note that, a pair of spiral waves with identical rotating chirality will not annihilate due to the general requirement of topological charge conservation (38). Remarkably, we find that the pair of opposite-charge spiral cores experience short-range attraction and intermediate-range repulsion for all nonreciprocity $\varepsilon > 0$; as the level of nonreciprocity increases, the attraction range shrinks while the repulsion range expands (Fig. 4B). As a consequence, opposite-charge spiral cores would tend to avoid getting too close to each other at larger nonreciprocity, thus maintaining stability during the development of spiral wave patterns.

We stress that the emergence of propagating spiral waves observed in pilus-driven bacterial living matter does not rely on spatial inhomogeneity, in contrast to spiral waves in chemical excitable media (39). Nonetheless, we sought to understand how spatial inhomogeneity or defects would perturb the wave dynamics. To examine this question, we created a hollow area with inactivated oscillators (representing a spatial defect inert to oscillator interactions) in the simulation (Methods). We found that the wave morphology was only affected very close to the defect (Fig. 4C; Movie 10), which is confirmed by laser ablation experiment (Movie 11; Methods). Therefore, the global spiral wave pattern is resilient to spatial inhomogeneity.

**Discussion**

In summary, we have discovered a novel spiral wave pattern in the bacterial world: Synchronization of type-IV pilus motility in bacterial living matter gives rise to a large-scale propagating spiral waves of tension force. Using numerical simulations of a locally-coupled oscillator model, we show that nonreciprocal coupling between the pilus motor activities underlies the emergence of the spiral wave pattern.

For many systems out of equilibrium, including living matter and synthetic active matter, the interaction between their elementary units is often nonreciprocal (40-43), i.e. violating the action-reaction principle of Newton's Third Law.  Recently nonreciprocity has been suggested to drive phase transition and self-organization in active matter (43, 44).  In this connection, it is intriguing that nonreciprocity controls wave dynamics during the synchronization of coupled oscillators.  Our results may shed light on the collective dynamics in other living and active matter systems with cyclic force generating processes (45-47).

Most mechanisms for the onset of spiral waves require external stimulation (39), spatial inhomogeneity (39), or special initial conditions (6, 48).  By contrast, our results suggest that nonreciprocity provides a simple mechanism for the spontaneous formation of stable spiral waves in homogeneous media without external stimuli.  An important feature of the bacterial spiral wave pattern reported here is that the spiral cores are nearly stationary at steady state.  This characteristic is shared by certain types of electrical and chemical spiral waves found in oocyte cytoplasm (17), eukaryotic cell cortex (18), and cardiac tissues (7, 10).  The bacterial spiral wave pattern may therefore serve as a tractable mechanical analogue for investigating the origin and control of stable spiral waves in diverse living systems.

**Methods**

**Experimental procedures and data analysis**

The following strains were used: piliated *P. aeruginosa* PA14 ZK3367 (PA14 *flgK*::Tn5; without flagellar motility but retaining type-IV pilus motility; primarily used in this study), gift from Roberto Kolter, Harvard University (49); *P. aeruginosa* PA14 ZK3367 labeled by GFP [PA14 *flgK*::Tn5 transformed with a plasmid pMHLB P*lasB*-gfp(ASV) Gen[R] (50) by electroporation; the plasmid was a gift from Liang Yang at Southern University of Science and Technology]; *P. aeruginosa* PA14 ZK3367 with constitutive expression of mCherry in cytoplasm [PA14 *flgK*::Tn5 transformed with a pJN105-mCherry Gen[R] plasmid (gift from Fan Jin at Shenzhen Institute of Advanced Technology) by electroporation]; non-motile *P. aeruginosa* without either flagellar or type-IV pilus motility [PA14 *flgk::Tn5 ΔpilA*, which does not produce type IV pili; and PA14 *flgk::Tn5 ΔpilY1*, which is not able to retract and generate mechanical forces] (51), gifts from George A. O'Toole, Dartmouth College; *P. aeruginosa* PA14 ZK3367 deficient in PelA matrix production (PA14 *flgK*::Tn5 Δ*pelA*) or deficient in rhamnolipid production (PA14 *flgK*::Tn5 Δ*rhlA*) (52). Single-colony isolates were grown overnight (all for ~13-14 hr, except the strain harboring pJN105-mCherry plasmid that requires ~16 hr growth) in 10 mL culture tubes (unless otherwise stated) with shaking in LB medium (1% Bacto tryptone, 0.5% yeast extract, 0.5% NaCl) at 30 °C to stationary phase. For cultures of strains harbouring plasmids expressing GFP or mCherry, the antibiotic gentamycin was added to a final concentration 50 µg/mL in the LB broth. Overnight cultures were used for inoculating colonies on agar plates.

Colonies or disk-shaped artificial bacterial films of *P. aeruginosa* were grown on 0.5% Difco Bacto agar plates infused with M9DCAA medium (53) [20 mM $NH_4Cl$, 12 mM $Na_2HPO_4$, 22 mM $KH_2PO_4$, 8.6 mM NaCl, 1mM $MgSO_4$, 1mM $CaCl_2$, 11 mM dextrose, and 0.5% (wt/vol) casamino acids (BD Bacto, cat. No. 223050)]. As $Ca^{2+}$ cannot coexist stably with many ions, this medium was prepared and stored in two components: (1) 10X nutrient solution without $CaCl_2$, sterilized and stored at room temperature; (2) agar infused with $CaCl_2$ at $1\frac{1}{9}$ times of the desired concentrations, sterilized and stored in 100 mL aliquots. Before use, the component (2) was melted completely in a microwave oven and cooled to ~50-60 °C. For each plate, 18 mL molten component (2) was mixed with 2 mL component (1), and the mixture was poured to a polystyrene petri dish (90 mm diameter, 15 mm height). The M9DCAA agar plate was swirled gently to ensure surface flatness, followed by further drying under laminar airflow for 20 min at room temperature.

The detailed procedures of sample preparation, sample manipulation, image acquisition, image processing and data analysis, and rheological measurement are included in Supplementary Methods. In addition to the propagating spiral waves observed in piliated *P. aeruginosa* PA14 (PA14 *flgK*::Tn5) that was primarily used in this study, similar spiral wave pattern was observed with a piliated *P. aeruginosa* PA14 mutant lacking the major extracellular matrix component Pel (54) (PA14 *flgK*::Tn5 Δ*pelA*). The spiral waves were also observed with a rhamnolipid-deficient piliated *P. aeruginosa* PA14 mutant (PA14 *flgK*::Tn5 Δ*rhlA*), thus excluding the contribution of interfacial flows to the phenomenon.

**Theoretical analysis of displacement and tension fields in the bacterial film**

**(1) General relation between spatial and temporal derivatives in propagating waves.**
Without loss of generality, we define positive $x$-axis direction as the local wave propagating direction. Any mechanical variable $\Gamma$ (e.g., the displacement, speed, tension, surface packing cell density, oscillation phase angle, etc.) that periodically change in time can be described as a function of phase angle $\theta$ (in the range of $[0, 2\pi]$). As $\theta(x,t)$ can be expressed in terms of $(x - v_0 t)$ as $\theta_{(x-v_0 t)}$, where $v_0$ is the wave speed and it is assumed to be a constant for a stable periodically propagating wave, we have:

$$\Gamma(x,t) = \Gamma(\theta) = \Gamma(\theta_{(x-v_0 t)}). \quad \#(2)$$

Then we have:

$$\frac{\partial \Gamma}{\partial x} = \frac{d\Gamma}{d\theta}\frac{\partial \theta}{\partial x} = \frac{d\Gamma}{d\theta}\frac{d\theta}{d(x-v_0 t)}\frac{\partial(x-v_0 t)}{\partial x} = \frac{d\Gamma}{d\theta}\frac{d\theta}{d(x-v_0 t)},$$

$$\frac{\partial \Gamma}{\partial t} = \frac{d\Gamma}{d\theta}\frac{\partial \theta}{\partial t} = \frac{d\Gamma}{d\theta}\frac{d\theta}{d(x-v_0 t)}\frac{\partial(x-v_0 t)}{\partial t} = -v_0\frac{d\Gamma}{d\theta}\frac{d\theta}{d(x-v_0 t)}. \quad \#(3)$$

Comparing the form of $\frac{\partial \Gamma}{\partial x}$ and $\frac{\partial \Gamma}{\partial t}$, we have:

$$\frac{\partial \Gamma}{\partial x} = -\frac{1}{v_0}\frac{\partial \Gamma}{\partial t}. \quad \#(4)$$

Thus, we can replace the operator $\frac{\partial}{\partial x}$ by $-\frac{1}{v_0}\frac{\partial}{\partial t}$ in the following calculation. Meanwhile, since we have chosen the positive x-axis direction as the propagating direction of the wave, the y-axis direction is parallel to the wavefront where the oscillation phases $\theta$ are in sync. Thus, $\Gamma$ should not vary in y-axis direction and we have: $\frac{\partial \Gamma}{\partial y} = 0$.

**(2) Displacement field analysis**. The total biomass in a mass element of the quasi-2D bacterial film with surface area $S_0$ and surface cell density $\rho_0$ is proportional to $\rho_0 S_0$. Due to biomass conservation, we have the following relation between small variations in the surface area ($\Delta S \equiv S - S_0$) and cell density ($\Delta \rho \equiv \rho - \rho_0$):

$$\Delta \rho S_0 + \Delta S \rho_0 = 0. \quad \#(5)$$

We consider a small region in a bacterial film with propagating mechanical waves that approximate plane waves. The in-plane deformation of the small region can be described by the gradients of the displacement vector field $\vec{d}(x,y,t)$ of mass elements in the region, with the Cartesian components of this vector field $d_1(x,y,t)$ and $d_2(x,y,t)$ representing the displacement along x and y axis.

For a mass element located at $(x, y)$ with area $S_0$, when there is a displacement field $\vec{d}(x,y,t)$, the area of the mass element will be changed by:

$$\Delta S = S_0(\partial_x d_1 + \partial_y d_2) = S_0 \nabla \cdot \vec{d}. \quad \#(6)$$

Thus, $\Delta \rho$, the cell density variation associated with the deformation of the mass element, can be written as:

$$\Delta \rho = -\frac{\rho_0}{S_0}\Delta S = -\rho_0 \nabla \cdot \vec{d}. \quad \#(7)$$

In 1D, substituting $\partial_x$ with $-\frac{1}{v_0}\partial_t$, we have:

$$\Delta \rho = \frac{\rho_0}{v_0}\partial_t d_1. \quad \#(8)$$

Therefore, the cell density variation $\Delta \rho$ should be linearly proportional to the time rate of cell displacement $\partial_t d_1$, as shown in Fig. 2B (right panel).

**(3) Tension analysis**. The deformation of mass elements in the quasi-2D bacterial film is determined by local tension due to pilus activities, film elasticity, and substrate friction. The mechanical equilibrium satisfies:

$$0 = -\partial_i(P_0 - T) + \partial_j C_{ijkl}\partial_l d_k + \frac{f_i}{H}, \quad \#(9)$$

where $P_0$ is a constant pressure; $T$ is the tension field (which acts isotopically and thus can be regarded as a negative pressure); $C_{ijkl}$ is the elasticity tensor of the bacterial film; $d_k$ ($k = 1,2$) is the $k$-th Cartesian component of the displacement field $\vec{d}(x,y,t)$; $f_i$ is projection of friction force per unit area along the $i$-th axis; and $H$ is the height of bacterial film. The magnitude of friction force per area is assumed to be constant. Choosing $x$-axis as the wave propagation direction, we can ignore $d_2$ and simplify Eq. (9) to:

$$\partial_x T + E\partial_x^2 d_1 + \frac{f_1}{H} = 0, \quad \#(10)$$

Here $E = C_{1111}$ is Young's modulus. Making use of Eq. (4), we can substitute all the $\partial_x$ with $-\frac{1}{v_0}\partial_t$ and obtain the following equation:

$$0 = -\frac{1}{v_0}\partial_t T + \frac{E}{v_0^2}\partial_t^2 d_1 + \frac{f_1}{H}. \quad \#(11)$$

The tension $T$ can then be solved as follows:

$$T - T_0 = \frac{C}{v_0}\partial_t d_1 + v_0 \int_0^t \left(\frac{f_1}{H}\right) dt', \quad \#(12)$$

where $T_0$ is a constant. Combing Eq. (8) and Eq. (12), we have:

$$\Delta T \equiv T - T_0 = C\frac{\Delta\rho}{\rho_0} + v_0 \int_0^t \left(\frac{f_1}{H}\right) dt'. \quad \#(13)$$

According to Eq. (13), the variation in tension ($\Delta T$) is proportional to the variation in local cell density ($\Delta\rho$) plus a contribution from the friction $\int_0^t \left(\frac{f_1}{H}\right) dt'$. As the friction force has a constant magnitude but its sign depends on the direction of velocity, its time integral is a periodic function having the same phase as the displacement rate $\partial_t d_1$ or local cell density $\Delta\rho$. Therefore, the tension is a periodic function in phase with local cell density; in particular, the tension at the leading side of the wavefront is expected to be higher than that at the lagging side of the wavefront, because the cell density is higher at the leading side (see caption of Fig. 2A).

**Nonreciprocal coupled-oscillator model**

Our physical model incorporates both oscillator synchronization and continuum mechanics in the bacterial film. We modeled mass elements in the bacterial film as a system of spatially coupled phase oscillators on a two-dimensional square lattice. The phase angle of the oscillator residing in a mass element varies periodically in time while being modified by the phase angles of neighboring oscillators within the coupling range $l$ via nonreciprocal coupling (Eq. 1 of main text). The nonreciprocal coupling is enforced by the step function $F(x)$ in Eq. 1. For example, consider an arbitrary pair of interacting oscillators located at positions $(i,j)$ and $(m,n)$, respectively. Assuming that $\theta_{(i,j)}$ lags behind $\theta_{(m,n)}$ (i.e., $\theta_{(m,n)} > \theta_{(i,j)}$) without loss of generality, $\theta_{(i,j)}$ will be accelerated by $\sim |\sin(\theta_{(m,n)} - \theta_{(i,j)})| \cdot (1+\varepsilon)/2$ and conversely, $\theta_{(m,n)}$ will be decelerated by $\sim |\sin(\theta_{(i,j)} - \theta_{(m,n)})| \cdot (1-\varepsilon)/2$. The step function $F(x)$ ensures that the magnitude of the phase acceleration ($F > 0.5$) is higher than that of the phase deceleration ($F < 0.5$) during pair-wise interactions.

In addition to the oscillator synchronization dynamics, the position of mass elements is subject to displacement due to external forces, and the displacement of mass elements may change the number of neighbours within the coupling range $l$ for each oscillator. According to the definition of the phase angle $\theta$ of the oscillator residing in a mass element, the tension stress generated by a mass element located at $(x,y)$ varies periodically in time and can be written as $T(x,y,t) = b(\theta)T_{max}$, where $b(\theta) \in [0,1]$ is the duty ratio controlling the tension strength variation in the pilus retraction-extension cycle and the constant parameter $T_{max}$ is the maximum tension stress that can be generated by a mass element in the bacterial film. The simplest form of $b(\theta)$ is $b(\theta) = \frac{\theta}{2\pi}$, so we choose $T(x,y,t) = \frac{\theta}{2\pi}T_{max}$, where $\theta$ is in the range $[0, 2\pi]$.

While the phase angle distribution $\theta(x,y,t)$ evolves according to the non-reciprocal Kuramoto model (Eq. 1 in main text), the displacement field $\vec{d}(x,y,t) = (d_1, d_2)$ of mass elements in the bacterial film evolves according to the following force-balance equations based on Eq. [9]:

$$\partial_x T + E\partial_x^2 d_1 + G\partial_y^2 d_1 + f_1/H = 0,$$
$$\partial_y T + E\partial_y^2 d_2 + G\partial_x^2 d_2 + f_2/H = 0, \quad \#(14)$$

where $d_1$ and $d_2$ are the projection of the displacement field $\vec{d}(x,y,t)$ along $x$-axis and $y$-axis, respectively; $E$ is Young's modulus; $G$ is the shear modulus and it is related to Young's modulus in the form of $E = 2(1+v)G$, with $v$ being Poisson's ratio; $f_1$ and $f_2$ are projection of friction force per unit area along $x$-axis and $y$-axis, respectively; and $H$ is the height of bacterial film. The magnitude of friction force per area is assumed to be constant. The tension field $T(x,y,t) = \frac{\theta}{2\pi}T_{max}$ is obtained from the time evolution of $\theta(x,y,t)$. Finally, the cell density distribution $\rho(x,y,t)$ can be obtained from the displacement field as $\rho(x,y,t) = \rho_0(1 - \nabla \cdot \vec{d})$, where $\rho_0$ is the cell density in the absence of pilus-driven tension force (Eq. 7).

The simulations were performed in a 100x100 coupled oscillator system arranged on a square-lattice domain for all results unless otherwise stated. Oscillators in the simulation domain only interacted with neighbors within the distance of coupling length; therefore, the oscillators near the boundary interacted with fewer neighbors than those inside the domain. To account for the intrinsic noise in pilus activity of cells in experiments, the oscillators were gradually activated (i.e., being able to interact with neighbors according to Eq. 1) at the beginning the simulation, with the waiting time for activating each oscillator following a normal distribution (mean: 500 time steps, standard deviation: 167 time steps). The phase angle of each oscillator at the first time step when it was activated was randomly drawn from a uniform distribution in $[0, 2\pi)$. The intrinsic angular frequency of each oscillator $\omega_{0(i,j)}$ was drawn from a normal distribution (mean: $\bar{\omega}$, standard deviation: $\Delta\omega$). The Gaussian white noise $\eta(t)$ satisfies $\langle \eta(t)\eta(t')\rangle = \bar{\eta}^2 \delta(t-t')$, where $\bar{\eta}$ is the noise strength. The evolution of the phase angle of each oscillator was calculated in each time step by the difference equation of Eq. 1: $\Delta\theta_{(i,j)} = \Delta t \cdot \{\omega_{0(i,j)} + \frac{\kappa}{N}\sum_{r\leq l}[\sin(\theta_{(m,n)} - \theta_{(i,j)}) \cdot F(\sin(\theta_{(m,n)} - \theta_{(i,j)}))] + \eta_{(t)}\}$, where $\Delta t$ is the time step. The tension field at the next time step was then updated as $T(t + \Delta t) = \frac{\theta(t+\Delta t)}{2\pi} T_{max}$; with $T(t + \Delta t)$, Eq. 14 was solved in Fourier space to yield $\vec{d}(x, y, t + \Delta t)$, which was subsequently used in further evolving the phase angle distribution.

The simulation parameters used in Eq. 1 were chosen as follows: the intrinsic angular frequency of the oscillators $\omega_{0_{(i,j)}}$ was sampled from a Gaussian distribution with mean $\bar{\omega} = \pi/240$ rad/s and standard deviation $\Delta\omega = \pi/2400$ rad/s; nearest oscillator distance $d$, 1 pixel (corresponding to $40\ \mu m$ in physical space; i.e. mass element size being $40\ \mu m \times 40\ \mu m$); coupling length $l = 4d$, i.e., 4 pixels or $160\ \mu m$; coupling strength $\kappa$, $\pi/80$ rad/s; time step $\Delta t$, 0.1 (for Fig. 4 and Movie 10; corresponding to 8 s in experimental time) or 0.025 (for all other simulations; corresponding to 2 s in experimental time); strength of the Gaussian white noise $\bar{\eta}$, $\pi\sqrt{\Delta t}/5$ rad/s$^{\frac{3}{2}}$ (in all simulations except for Fig. 4B and 4C, where $\bar{\eta}$ was set as 0 to ensure that the spiral wave pattern was stable). We found that the coupling length controls the wavelength of the resultant spiral wave pattern (estimated as ~ $2\pi/|\nabla\theta(x,y,t)|$) in the simulations; the value of coupling length $l = 160\ \mu m$ was chosen so as to produce a wavelength comparable to that found in experiments. For Eq. 14, $T_{max}$, the maximum tension stress (tension force per unit area) generated by a mass element in the bacterial film was estimated as $T_{max} = (F_{pilus}/S_{cell}) \cdot n_{pili} \cdot \chi = 81$ Pa; here $F_{pilus}$ is the retraction force of one single pilus, ~100 pN (15); $S_{cell} = [\frac{4}{3}\pi a^3 + \pi a^2(l_B - 2a)]^{2/3}$ is the surface area of a cell, with $l_B = 3\ \mu m$ and $a = 0.4\ \mu m$ being the length and radius of a rod-shape bacterial cell; $n_{pili} = 4$ is the approximate number of Type-IV pili on one cell (20); and $\chi$ is the fraction of pilus-generated tension stress that applied along an arbitrary direction, which can be calculated as: $\chi = \int_0^{\frac{\pi}{2}} \int_{-\pi}^{\pi} \sin\phi \cos\phi \, d\phi d\psi / \int_0^{\pi} \int_{-\pi}^{\pi} \sin\phi \, d\phi d\psi = 1/4$, where $\phi$ is the zenith angle in spherical coordinate system (assuming that bacterial pili apply isotropic retraction forces in all directions). The shear modulus $G = 40$ Pa was chosen according to our rheological measurement of artificial bacterial films (see previous section). Assuming the Poisson's ratio of the bacterial film to be 0.4, the Young's modulus $E$ was estimated as $E = 2(1 + \nu)G = 112$ Pa. The friction force per unit area $f$ should be smaller than $T_{max}$, otherwise isolated cells would not be able to move on a substrate, so $f$ was chosen to range from 1 to 10 Pa; note that $f$ suppresses displacement but does not affect wave dynamics. Periodic boundary condition was used in the numerical simulations combining Eq. 1 and Eq. 14.

To study the effect of forced synchronization of oscillators in an area on the dynamics of spiral wave development (Fig. S5, Movie 8; i.e. to simulate the treatment of the artificial bacterial films with drugs that trigger pilus retraction), we forced the oscillators along one side of the boundary of the square simulation domain to synchronize their phases by manually setting the phase angle of these oscillators to be a linear function of time $\theta(t) = \omega_0 t$, where $\omega_0 = \frac{\pi}{160}\ rad/s$ is a constant.

To study the dynamics of spiral core interactions (Fig. 4B), the displacement of mass elements was neglected and the simulations were done with Eq. 1 alone with vanishing boundary condition. We artificially created two counter-rotating spiral wave patterns with opposite rotating chirality that occupied different halves of the simulation domain, by initializing the phase angle distribution of oscillators in one half as an Archimedean spiral $\theta(\rho, \phi) = \phi - \frac{\rho}{k}$ (where $(\rho, \phi)$ are polar coordinates and $k = 4d$ is a constant) and in the other half as the mirror-symmetric form of $\theta(\rho, \phi)$. The noise term in Eq. 1 was turned off and the intrinsic angular frequency of oscillators was set as a constant, in order to exclude the effect of randomness on spiral core interactions.

To study the effect of spatial inhomogeneity or defects on a spiral wave with well-defined morphology and spiral core position (Fig. 4C; Movie 10, upper two rows), a system of 200x200 coupled oscillators was used; the displacement of mass elements was neglected and the simulations were done with Eq. 1 alone with vanishing boundary condition. We first artificially created a spiral wave with well-defined morphology and spiral core position by initializing the phase angle distribution of oscillators as an Archimedean spiral described above, with the initial spatial pattern of the wavefront (where the phase angle is zero) being $\rho = k\phi$. The system was allowed to evolve until reaching steady state, when the system displayed a stable propagating spiral wave pattern with the spiral core located at the center of the simulation domain. Then an area with a radius of $12d$ and with its center located at a distance $32d$ from the spiral core was chosen as the spatial defect. Oscillators in the hollow-like defect area were inactivated. The noise term in Eq. 1 was turned off and the intrinsic angular frequency of oscillators was set as a constant, in order to exclude the effect of randomness on the interaction between spatial defect and existing spiral waves.

To study the effect of spatial inhomogeneity or defects on spontaneously developed spiral waves (Movie 10, lower two rows), a system of coupled oscillators was allowed to evolve spontaneously until reaching steady state, when the system displayed stable propagating spiral waves with a variable number of spiral cores in the simulation domain. Then oscillators in an area with a radius of $25d$ and located at the center of the simulation domain were inactivated (Movie 10, lower two rows).

**Supplementary Materials**, including Supplementary Videos, is available in the online version of the paper.

**Data availability.** The data supporting the findings of this study are included within the paper and its Supplementary Materials.

**Code availability.** The custom codes used in this study are available from the corresponding author upon request.

**Acknowledgements**. We thank Fan Jin (Shenzhen Institute of Advanced Technology), Roberto Kolter (Harvard University), George A. O'Toole (Dartmouth College) and Liang Yang (Southern University of Science and Technology) for generous gifts of bacterial strains. We also thank Qi Ouyang, Alexandre Persat, Steven Strogatz, Leihan Tang and Zhigang Zheng for helpful discussions. This work was supported by the Ministry of Science and Technology Most China (No. 2021YFA0910700), National Natural Science Foundation of China (NSFC No. 31971182), and the Research Grants Council of Hong Kong SAR (RGC Ref. No. 14306820, 14307821, RFS2021-4S04 and CUHK Direct Grants). Y.W. acknowledges support from New Cornerstone Science Foundation through the Xplorer Prize.

**Author Contributions**: S.L. designed the study, performed experiments, developed the model, performed simulations, analyzed and interpreted the data. Y.L. made the initial observation. Y. Wang designed the laser ablation setup. Y. W. conceived the project, designed the study, analyzed and interpreted the data. Y.W. wrote the paper with S.L.'s input.

**Author Information**: Reprints and permissions information is available at the journal website. The authors declare no competing financial interests. Requests for materials should be addressed to Y.W. (ylwu@cuhk.edu.hk).

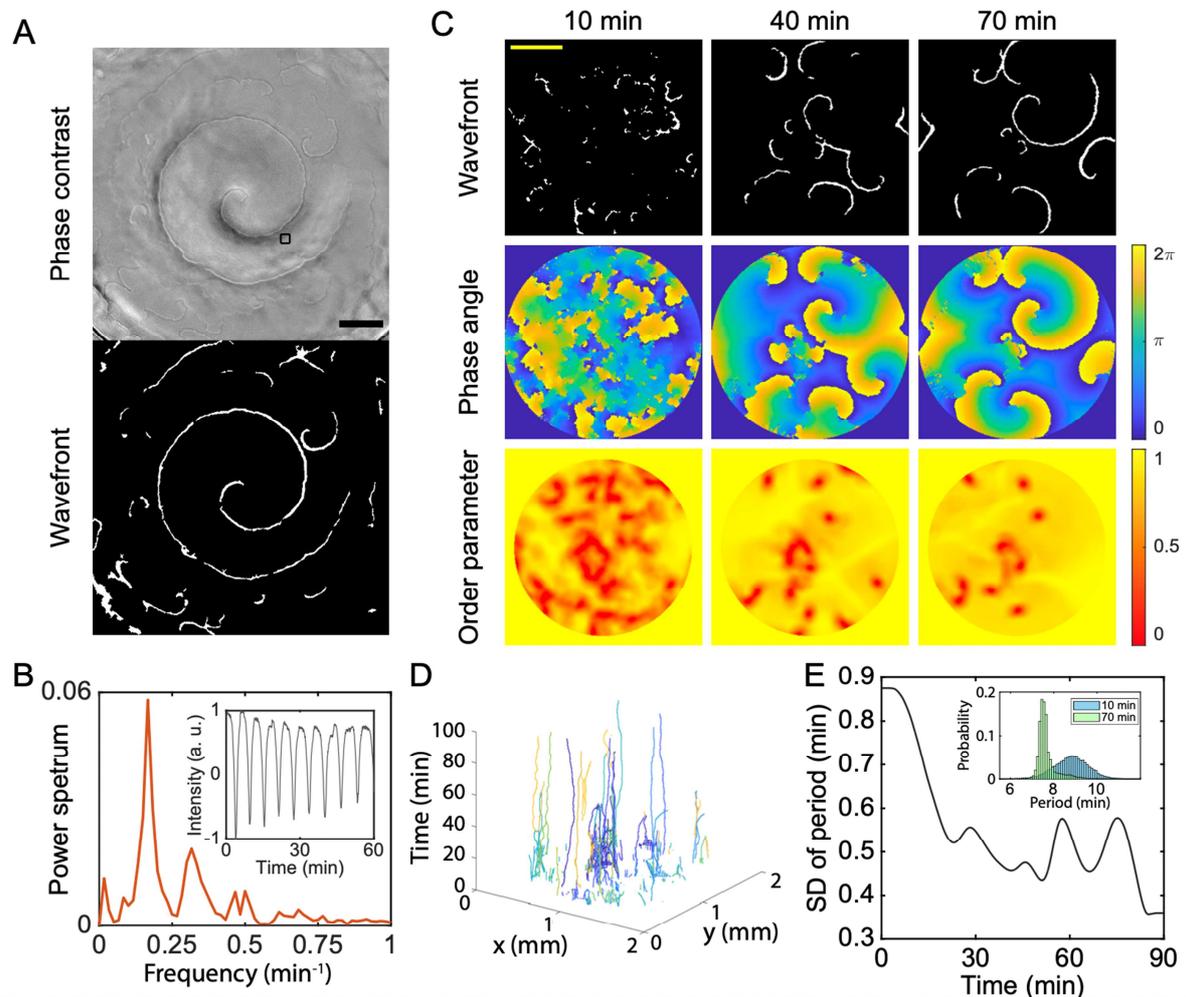

**Fig. 1. Self-organization of propagating spiral waves with stationary cores in quasi-2D bacterial living matter.** (A) Spiral wave pattern in a piliated *P. aeruginosa* (PA14 *flgK*::Tn5) colony. Wavefronts of the propagating spiral waves are visible in background-corrected phase contrast images (upper) and can be traced based on the variation of light intensity in a sequence of phase contrast images (lower) (Methods). Scale bar, 500 μm. Also see Movie 1 for the original phase contrast images. (B) Fourier-transformed power spectrum of the temporal variation of phase-contrast image intensity (inset) inside the black box specified in panel A (100.8 μm × 100.8 μm) (Methods). The power spectrum shows a prominent first-order peak at 0.17 min$^{-1}$. Scale bar, 500 μm. (C) Development of propagating spiral waves in artificial bacterial films of *P. aeruginosa*. Upper row: Traces of wavefronts (Methods). Middle row: Phase angle distributions of the waves (Methods). Lower row: Distributions of local order parameter calculated based on the phase distributions (Methods). Color bars to the right of the middle and lower rows indicate the magnitude of phase angle and local order parameter, respectively. Scale bar, 500 μm. Also see Movie 2. T= 0 min marks the onset of segmented and disordered wavefronts. (D) Space-time trajectories of spiral cores during the development of spiral waves associated with panel C. The nearly straight trajectories in this 3D plot at Time > ~20 min indicate that the spiral cores are almost stationary. (E) Standard deviation of the local oscillation period distribution (associated with spiral waves in

panel C) plotted against time in a representative experiment. Inset: Histograms showing the probability distribution of the local oscillation period at Time = 10 min and 70 min.

Figure 2

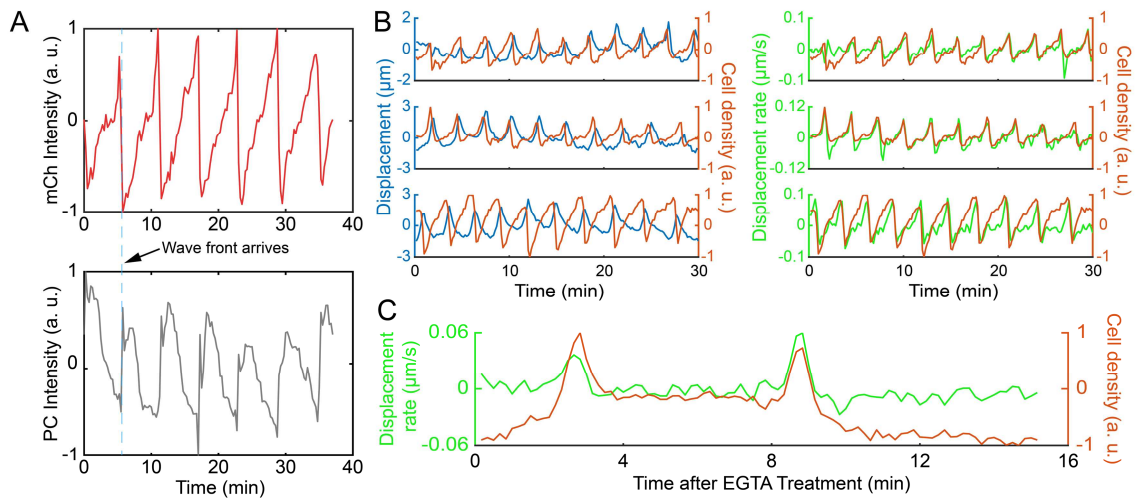

**Fig. 2. Analysis of cell density variation and single-cell displacement reveals propagating tension waves**. (A) Periodic variation of local cell density during spiral wave propagation. For a small region (~32.5 μm × 32.5 μm) arbitrarily selected from artificial bacterial films of mCherry-labeled *P. aeruginosa* that display propagating spiral waves, the fluorescence intensity of mCherry (mCh, upper panel) exhibits periodic oscillation coinciding with the light intensity oscillation in phase-contrast channel (lower panel). Dashed line indicates one of the instants when the wavefront arrives at the chosen position. The fluorescence intensity adjacent to the left of the dashed line corresponds to cell density just before the wavefront arrives, and thus it represents the cell density in front of the wavefront (denoted as leading side in main text) in a reference frame co-moving with the wavefront; conversely, the fluorescence intensity adjacent to the right of the dashed line represents the cell density behind the wavefront (denoted as lagging side). (B) Oscillatory forth-and-back displacement of individual cells and local cell density variation during spiral wave propagation. Traces measured with 3 representative cells from different experiments are presented. For each cell, the time trace of displacement (blue lines; left panel) or the rate of displacement (green lines; right panel) are overlaid by the time trace of local cell density (red lines) measured as the fluorescent intensity in the area associated with the cell (Methods). The time-averaged cross-correlation between displacement and cell density for the three traces in the left panel is: -0.03, 0.08 and 0.17, respectively (upper to lower); the cross-correlation between displacement rate and cell density for the three traces is 0.71, 0.86 and 0.71, respectively. (C) Forth-and-back displacement of individual cells and local cell density variation during the propagation of train waves triggered by EGTA treatment. The rate of displacement (green line) is overlaid by the time trace of local cell density (red line) measured as the fluorescent intensity in the area associated with the cell (Methods). The EGTA filter disk was applied to the bacterial film at T=0 min. In all panels the time traces with arbitrary unit are rescaled and vertically shifted, such that the values of any quantity over the entire time frame are in between +1 and -1.

Figure 3

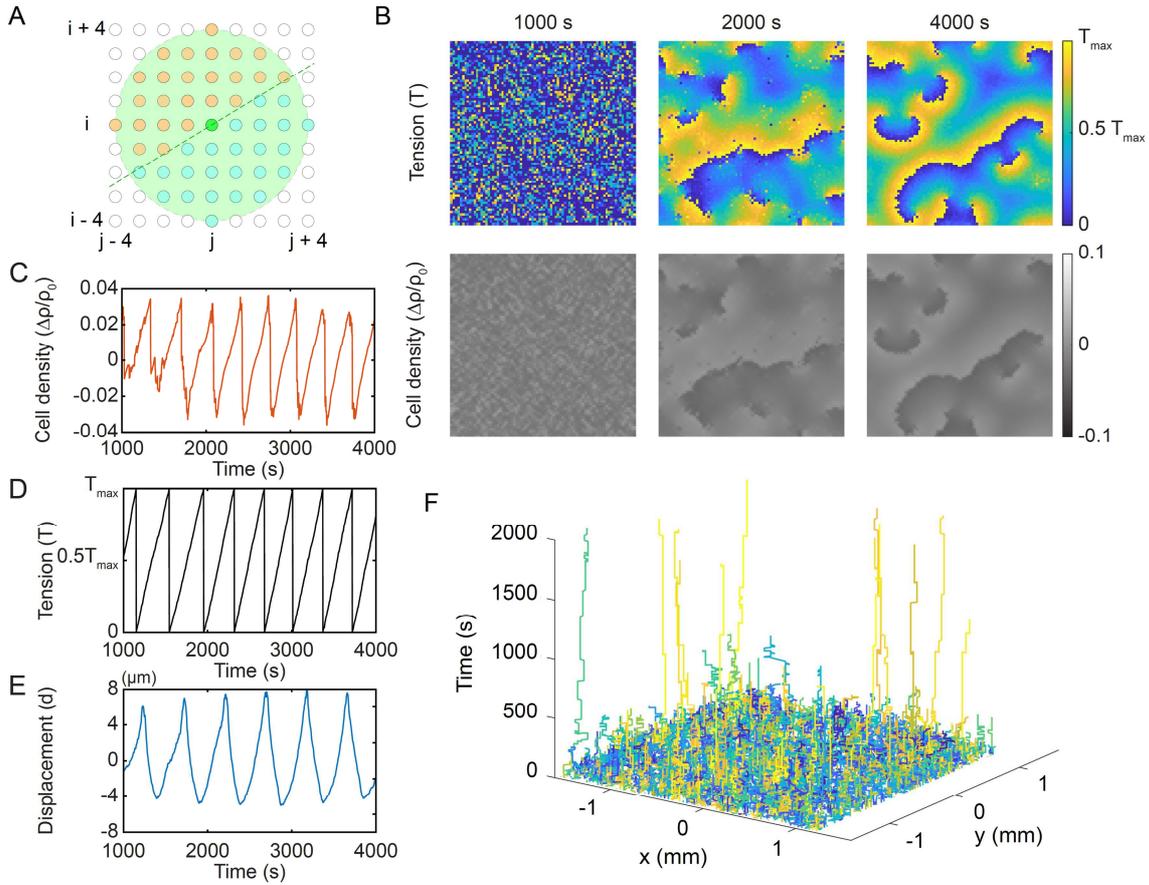

**Fig. 3. Coupled-oscillator model with nonreciprocal interaction reproduces the observed propagating spiral waves.** (A) Schematic diagram of the locally coupled two-dimensional Kuramoto model with isotropic nonreciprocal coupling. The nearest neighbor distance between the oscillators (represented as circles) is $d$. For the oscillator located at position $(i,j)$ at the center of the diagram, the green-shaded circular area with radius $l = 4d$ represents the coupling range. Assuming that the wavefront (i.e., equi-phase line indicated by the green dashed line) propagates towards the upper left corner, the oscillators labeled in blue color phase-lead oscillator $(i,j)$ and accelerate it, while those labeled in orange color lag behind oscillator $(i,j)$ and decelerate it. According to the summation in Eq. (1) of main text, the total magnitude of acceleration is greater than that of deceleration. (B) Spatiotemporal dynamics of tension field and cell density distribution in simulations with a full nonreciprocity $\varepsilon = 1$ showing the development of propagating spiral waves. The system consists of 80x80 oscillators. (C-E) Simulated temporal dynamics of cell density, tension and displacement at an arbitrarily selected mass element during spiral wave propagation $T_{max} = 81$ Pa in panel D (Methods). (F) Space-time trajectories of spiral cores during the development of spiral waves associated with panel B plotted in the same manner as Fig. 1D.

Figure 4

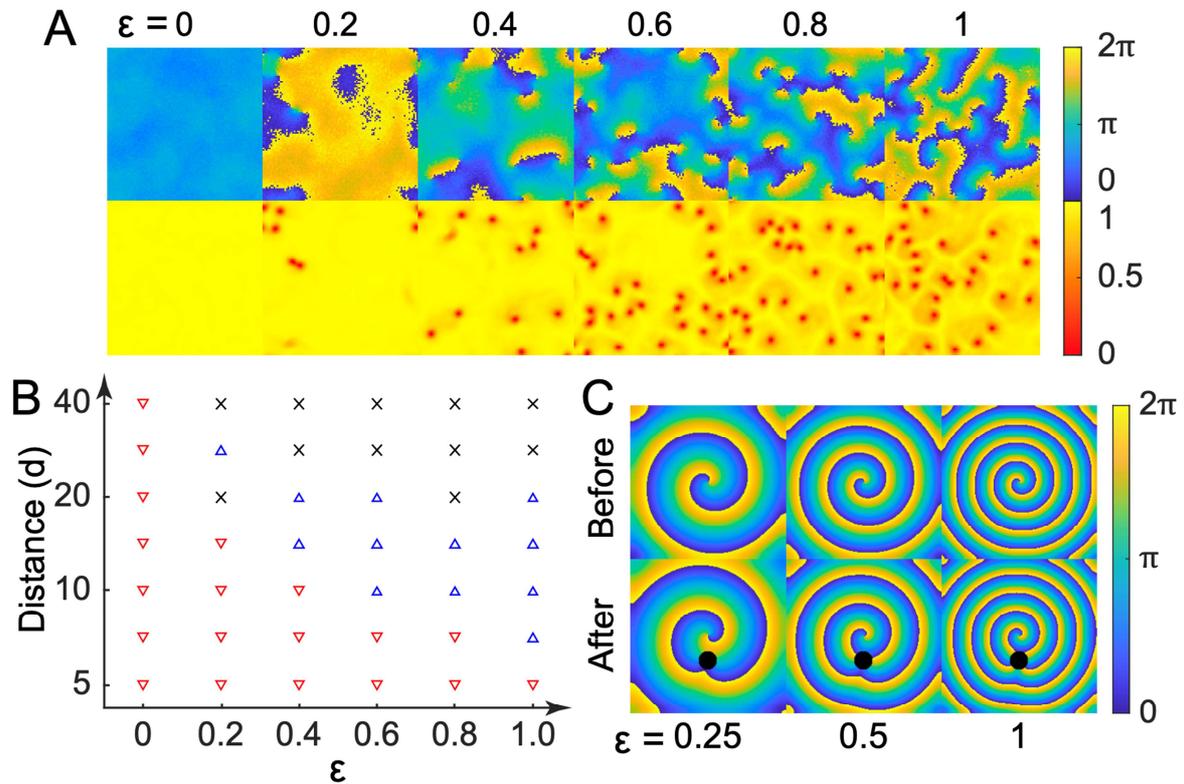

**Fig. 4. Effect of nonreciprocity and spatial inhomogeneity on the stability of spiral wave pattern**. (A) Spiral wave patterns formed with different levels of nonreciprocity in simulations of the coupled oscillator model. The spatial distributions of phase angle (upper) and local order parameter (lower) at Time = 4000 s are shown for the corresponding level of nonreciprocity $\varepsilon$ indicated on top of the figures. (B) Phase map of pairwise-interaction mode between spiral cores on the plane of nonreciprocity ($\varepsilon$) and core-core distance (in unit of nearest oscillator distance $d$). For a given $\varepsilon$, the pair of spiral cores may move towards each other ($\nabla$), repel from each other ($\Delta$), or display no relative motion (×), depending on their distance. (C) Resilience of spiral waves to spatial inhomogeneity. This panel shows the steady-state spatial distributions of phase angle before (upper row) and after (lower row) introducing a hollow-like spatial defect in the simulation domain (black solid circle), where the oscillators are inactivated and do not interact with others (Methods).

# Supplementary Information

## Supplementary Figures

Figure S1

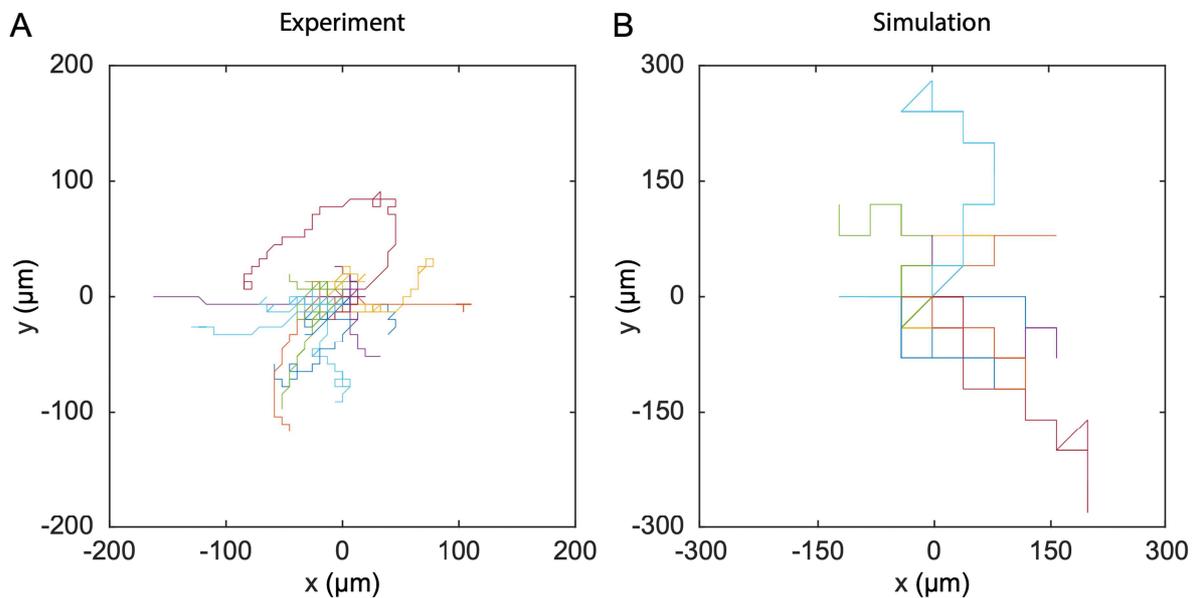

**Fig. S1. Trajectories of spiral cores of propagating spiral waves at steady state.** (A) Representative spiral core trajectories acquired experimentally in artificial bacterial films of piliated *P. aeruginosa* (PA14 *flgK*::Tn5). Each trajectory lasted for ~1 hour. The range of spiral core motion is a few tens of µm and two order of magnitude smaller than the wavelength (~1 mm). The trajectories appeared random and showed no periodic meandering motion. The trajectories of different cores are labeled with different color. The initial position of each core is set to x = 0, y = 0 for better comparison. (B) Representative spiral core trajectories in numerical simulations of the nonreciprocal coupled oscillator model with a full nonreciprocity ($\varepsilon = 1$) associated with main text Fig. 3F. The trajectories are plotted in the same manner as in panel A.

Figure S2

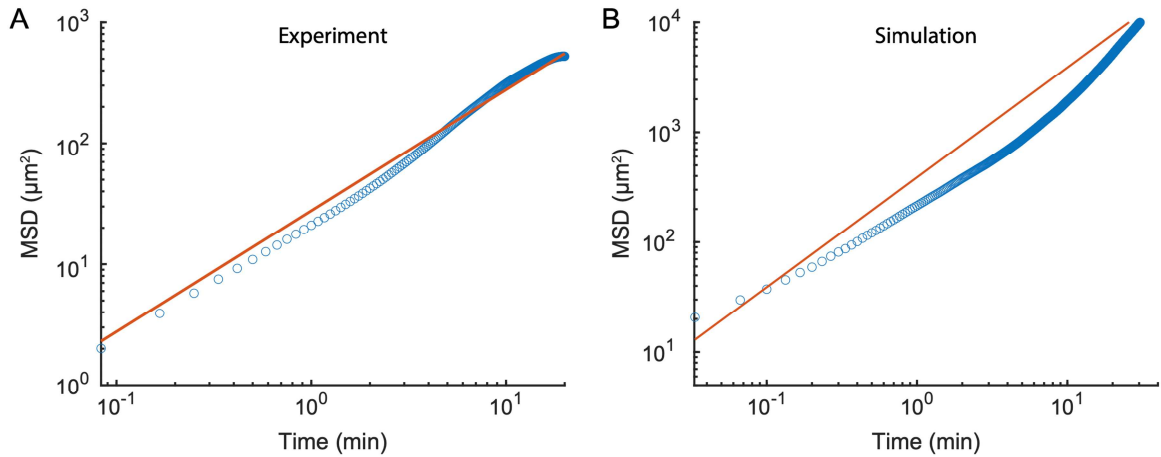

**Fig. S2. Mean-square displacement (MSD) and diffusivity of spiral cores measured in experiments (panel A; with 20 spiral cores) and in simulations (panel B; with 15 spiral cores).** Blue circles: MSD data. Red line: MSD fitted to $4Dt$, with $D$ being the diffusion coefficient. The collective MSD of all spiral cores was computed by segmenting spiral core trajectories with different time windows. Note that the trajectories of spiral cores obtained in experiments were corrected for their own drift, which was likely caused by cell growth during the tracking period; for the $i$-th core, its drift (denoted as $v_d$) was computed by fitting the core's own MSD (denoted as $MSD_i$) to $at + (v_d t)^2$, and its drift-corrected MSD is calculated as $\widetilde{MSD}_i = MSD_i - (v_d t)^2$. Then the overall drift-corrected MSD of $N$ spiral cores, defined as $MSD_{all} = \sum \widetilde{MSD}_i / N$, was fitted to $MSD_{all} = 4Dt$ as shown in panel A. In panel A, $D = 6.92 \ \mu m^2$/min (equivalent to $4.80 \times 10^{-5}$ square wavelengths per period). In panel B, $D = 9.72 \times 10^1 \ \mu m^2$/min (equivalent to $6.14 \times 10^{-4}$ square wavelengths per period).

Figure S3

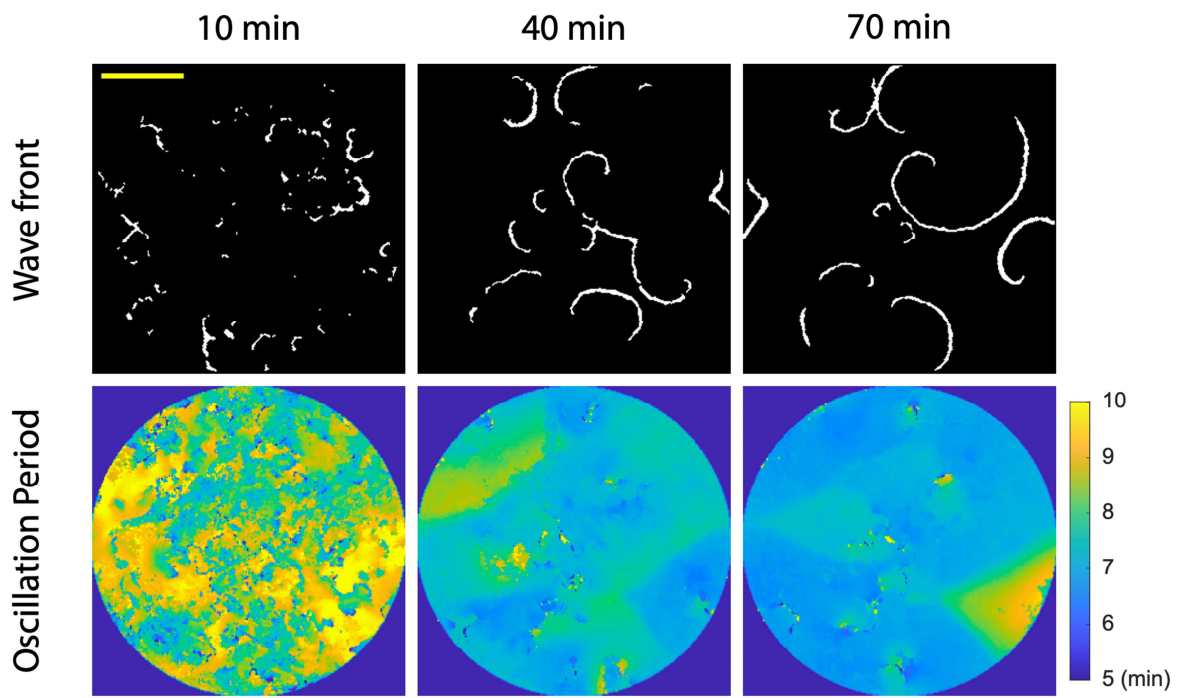

**Fig. S3. Development of propagating spiral waves in disk-shaped artificial bacterial films of *P. aeruginosa*.** Upper row: Traces of wavefronts (identical to upper row of main text Fig. 1C). Lower row: Distributions of local instantaneous oscillation period during the spiral wave development (Methods). Color bar to the right of the lower row indicates the magnitude of instantaneous oscillation period (unit: min). Scale bar, 500 μm. Also see Movie 2.

Figure S4

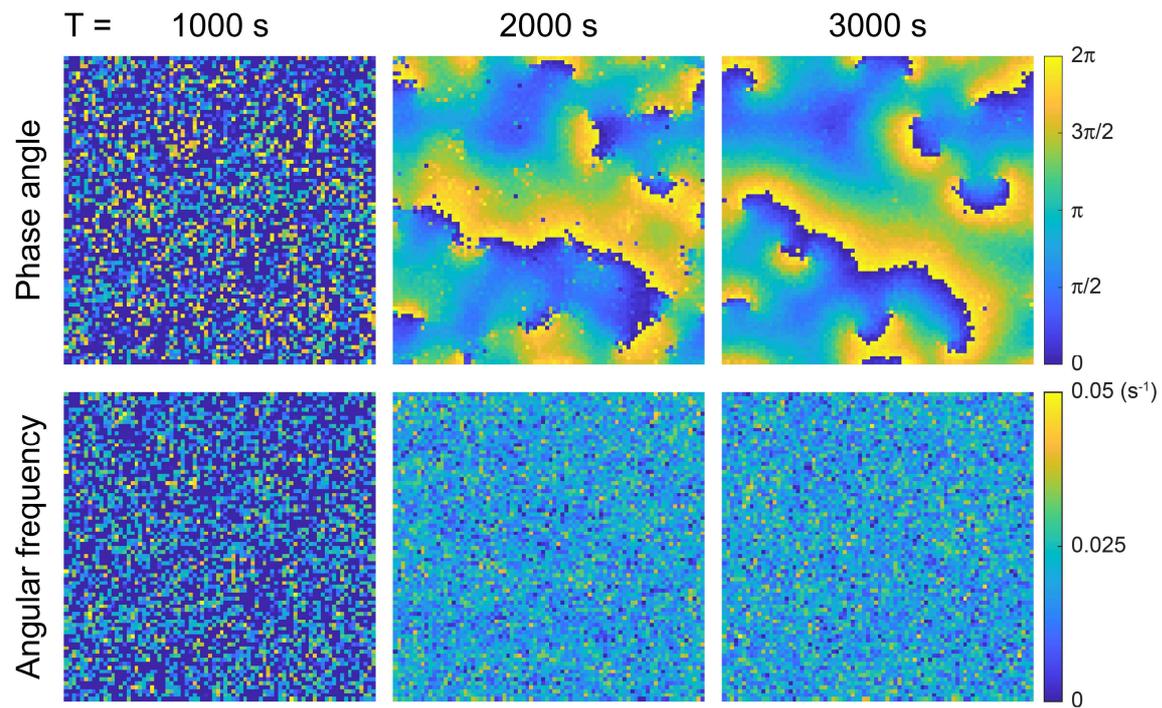

**Fig. S4. Spatiotemporal dynamics of phase angle (upper) and angular frequency distribution (lower) in simulations of the nonreciprocal coupled oscillator model with a full nonreciprocity $\varepsilon = 1$.** Upper row: Phase angle distribution (identical to main text Fig. 3B). Lower row: Distributions of local instantaneous angular frequency during the spiral wave development (Methods). Color bar to the right of the lower row indicates the magnitude of instantaneous oscillation angular frequency. Also see Movie 7.

Figure S5

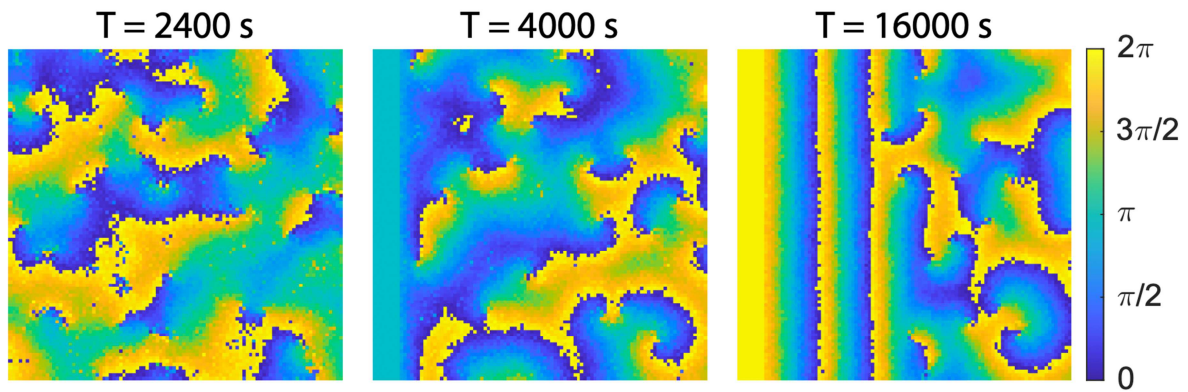

**Fig. S5. Generation of train waves by forced synchronization of oscillators in the nonreciprocal coupled oscillator model with a full nonreciprocity $\varepsilon = 1$.** The simulation was performed in a system with pre-existing propagating spiral waves (left panel, T= 2400 s). Starting from T= 2408 s, the oscillators near the left boundary were forced to synchronize and remained synchronized throughout the rest of the simulation, which simulated the application of a filter disk infused with drugs that trigger pilus retraction. Following the forced synchronization, train waves emanated from near the left boundary and propagated along a direction perpendicular to the boundary (right panel, T = 4000 s and T = 16000 s). The train waves gradually replaced the existing spiral waves. Also see Methods and Movie 8. The results reproduce the experimental phenomena in Movie 5 and Movie 6, supporting the notion that train waves observed in the experiments were due to forced synchronization of the pilus retraction-extension cycle.

**Legends of supplementary movies**

Movie 1. Propagating spiral waves in a naturally developed piliated *P. aeruginosa* (PA14 *flgK*::Tn5) colony. This video is associated with main text Fig. 1A. Upper: Time-lapse phase contrast images. Lower: Wavefronts of the propagating spiral waves (labeled as white traces) traced by the variation of light intensity in the phase contrast images. The elapsed time is indicated by the time stamp (format: hh:mm:ss). Scale bar, 500 µm.

Movie 2. Onset and evolution of the propagating spiral waves in an artificial bacterial films of *P. aeruginosa* 14. This video is associated with main text Fig. 1C. Left: Time-lapse phase contrast images. Middle: Traces of wavefronts. Right: Phase angle distributions of the waves. Color bar to the right of the right panel indicate the magnitude of phase angle. The elapsed time is indicated by the time stamp (format: hh:mm:ss). Scale bar, 500 µm.

Movie 3. Local cell density variation during the propagation of spiral waves. This video is associated with main text Fig. 2A. All cells in the artificial bacterial film were labeled by mCherry fluorescence protein. Left: Time-lapse phase contrast images. Right: Time-lapse mCherry fluorescence of the bacterial film (as a proxy of cell density). The mCherry fluorescence intensity exhibits the same wave pattern that coincides with the wave observed in phase-contrast images. The elapsed time is indicated by the time stamp (format: hh:mm:ss). Scale bar, 500 µm.

Movie 4. Periodic forth-and-back displacement of cells during the propagation of spiral waves. This video is associated with main text Fig. 2B. Cells in the artificial bacterial film were labeled by either GFP (99.9%) or mCherry (0.1%) fluorescence protein. Left: Time-lapse video of GFP fluorescence of the bacterial film (as a proxy of cell density). Right: Time-lapse fluorescence video of mCherry-labeled cells in the bacterial film showing the periodic forth-and-back displacement of cells. The elapsed time is indicated by the time stamp (format: hh:mm:ss). Scale bar, 500 µm.

Movie 5. Effect of EGTA on the propagating spiral waves. The time-lapse phase contrast videos show that, when a filter paper disc soaked with EGTA was placed onto an artificial bacterial film with (right panel) or without (left panel) pre-existing spiral waves, a train of waves emanated from near the edge of the filter paper disc and propagated outwards in both conditions. The elapsed time is indicated by the time stamp (format: hh:mm:ss). Scale bar, 500 µm.

Movie 6. Effect of filter disks soaked with phenothiazines and ATP on the propagating spiral waves. The time-lapse phase contrast videos show that, when a filter paper disc soaked with trifluoperazine, thioridazine or ATP solutions (Methods) was placed onto an artificial bacterial film, a train of waves emanated from near the edge of the filter paper disc and propagated outward. By contrast, empty filter disks that were soaked with DI water or NaCl or sucrose solutions (Methods) cannot trigger train waves when placed on an artificial bacterial film (right panel). The elapsed time is indicated by the time stamp (format: hh:mm:ss). Scale bars, 500 µm.

Movie 7. Simulation results of the physical model incorporating both Kuramoto dynamics and continuum mechanics in the bacterial film. The video is associated with Fig. 3 and shows the spatial distribution of tension (left) and cell density variation (right) during the development of spiral waves. The simulation was performed in an 80x80 coupled oscillator system.

Movie 8. Generation of train waves by forced synchronization of oscillators in the simulation. The video is associated with Fig. S5. The simulation was performed in a 100x100 coupled oscillator system. At T = 2000 s, propagating spiral waves had already stabilized and the oscillators near the left boundary were forced to synchronize, simulating the application of a filter disk infused with drugs that trigger pilus retraction (see Movies 5, 6). Train waves emanate from near the left boundary and propagated along a direction perpendicular to the boundary, gradually replacing the existing spiral waves.

Movie 9. Evolution of phase angle and local order parameter distributions in simulations of the nonreciprocal coupled oscillator model at different levels of nonreciprocity. The simulation was performed in a 100x100 coupled oscillator system. Upper: phase angle distribution at different levels of nonreciprocity; lower: local order parameter distribution corresponding to the phase angle distribution right above it, with the level of nonreciprocity indicated at the lower left corner. Starting from a disordered initial phase angle distribution, stable propagating spiral waves emerge at all levels of nonreciprocity; by contrast, transient spiral waves appeared in the case of zero nonreciprocity ($\varepsilon = 0$) and they eventually disappeared, leaving an almost homogeneous phase angle distribution. Simulation time steps elapsed for all panels are indicated in the lower left-most panel.

Movie 10. Effect of hollow-like spatial defect on the propagating spiral waves in the simulation. The level of nonreciprocity is indicated at the top of each simulation. Upper two rows: The videos are associated with main text Fig. 4C and show the spatial distributions of phase angle (first row) and local order parameter (second row) before and after introducing a defect (black solid circle; from T=1000 time steps to a simulation domain with a single pre-existing spiral core at the center (Methods). Oscillators in the defect were inactivated and did not interact with others. The simulation was performed in a 200x200 coupled oscillator system. The wave dynamics was only deformed very close to the defect and the spiral core (specified by the local minima of local order parameter field) did not move. Lower two rows: The videos show the spatial distributions of phase angle (third row) and local order parameter (fourth row) before and after introducing a defect to the system (black solid circle; from T=1000 time steps) in simulations with pre-existing propagating spiral waves (Methods). The simulation was performed in a 100x100 coupled oscillator system. The wave dynamics was largely unaffected by the defect, except that at relatively low $\varepsilon$ the spiral cores close to the defect tended to get attracted towards it and subsequently disappeared when entering the defect area.

Movie 11. Effect of hollow-like spatial defect on the propagating spiral waves in laser ablation experiment. Laser (~250 mW, 444 nm) was illuminated onto a small area of an artificial bacterial film with propagating spiral waves, which ablated cells in the illuminated area in 1 min and created a spatial defect in the area unable to mediate wave propagation. The time-lapse videos (left panel: phase contrast; right: phase angle distribution) recorded the wave dynamics before (before 22 min 45 s) and after (starting from 24 min 59 s) the creation of spatial defect by laser ablation. The videos show that the spatial defect only affected the wave dynamics very close to the defect. The elapsed time is indicated by the time stamp (format: hh:mm:ss). Scale bars, 500 µm.